# Triadic Novelty: A Structural Typology of Science Innovation

Jin Ai,[1]* Richard S. Steinberg,[2] Chao Guo[3], Filipi Nascimento Silva[4]

[1] School of Public Affairs and Administration, Rutgers University, Newark, NJ, USA. jin.ai@rutgers.edu

*Corresponding author*

[2] Department of Economics and Lilly Family School of Philanthropy, Indiana University Indianapolis, IN, USA. rsteinbe@iu.edu

[3] School of Social Policy and Practice, University of Pennsylvania, Philadelphia, PA, USA. chaoguo@sp2.upenn.edu

[4] Center for Science of Science & Innovation - Kellogg, Northwestern University, Evanston, IL, USA. filipi.nascimento@kellogg.northwestern.edu

## Abstract

Scientific progress depends on novelty, but current evaluation systems often conflate novelty with recognition, favoring work that aligns with existing paradigms over ideas that challenge them. We introduce a theory-driven framework that conceptualizes novelty as a structural process rather than a single scalar outcome. Drawing on network science and theories of scientific discovery, we develop a triadic typology of novelty: Pioneers introduce new topics, Mavericks recombine distant areas of knowledge, and Vanguards reinforce weak but emerging connections. We apply this framework to philanthropic and nonprofit studies, an interdisciplinary and evolving field well suited to examining how novelty is recognized before evaluation norms are fully stabilized. Results show that novelty is not uniformly rewarded: Pioneer contributions are often weakly recognized unless later taken up, Maverick contributions receive consistent recognition, and Vanguard contributions exhibit diminishing returns as ideas become established. By distinguishing how novelty enters and evolves within knowledge structures, the framework offers a citation-independent approach with implications for research evaluation, funding design, and science and innovation policy.

## Introduction

Some ideas reshape entire fields only after long periods of neglect. When Joseph Schumpeter introduced the notion of "*Creative Destruction"*, the concept initially challenged dominant economic thinking and gained little immediate traction [1]. Yet decades later, it became foundational to modern theories of innovation and economic growth. This pattern, early resistance followed by later recognition, is not anomalous. It reflects a fundamental feature of scientific development, that is, novelty is social constructed and often recognized after institutional, cognitive, and social conditions align [2–5]. Novel research frequently lacks legitimacy at the moment of its introduction and is therefore more likely to be ignored, undervalued, or even penalized, despite its potential to reshape knowledge fields [2,5–7]. Thomas Kuhn described this dilemma as the "essential tension" between paradigm-challenging innovation and the stabilizing forces of normal science [3]. This challenge has become more pressing as contemporary research crosses disciplinary boundaries spanning the natural sciences, engineering, social sciences, and humanities [8], where evaluative norms are less standardized and novelty is harder to recognize ex ante.

A growing literature has proposed quantitative metrics designed to capture scientific novelty. These measures generally fall into two categories: citation-based proxies and direct novelty metrics. Citation-based proxies infer novelty indirectly from later citation outcomes, treating highly cited or atypically cited work as innovative [9]. However, rather than measuring novelty itself, citations primarily reflect recognition [10,11], which is shaped by institutional visibility, disciplinary gatekeeping, and cumulative advantage, all of which may lag substantially behind the introduction of novel ideas [5,12,13]. As a result, citation-based measures privilege forms of novelty that are readily assimilated into existing intellectual structures, while undervaluing ideas that depart from dominant paradigms [5,13]. Timing further complicates the use of citation-based measures another limitation. Because they rely on finite forward windows, therefore disadvantage ideas whose influence emerges slowly or only after shifts in evaluation norms [5,12,14]. Empirical studies show that some highly original contributions received delayed recognition [14,15], and those with high citation impacts are not necessarily the most novel work perceived by the field [9]. This evidence suggests that citation-based proxies are insufficient for capturing novelty as it arises, rather than it is later recognized.

Direct novelty metrics seek to address this limitation through a variety of approaches. The ***first*** line focuses on atypical combinations, defining novelty as the rare co-occurrence of concepts, references, or topics that have not previously appeared together [5,13,16]. These approaches capture an important mechanism of innovation, recombination, but typically summarize novelty using a continuous deviation score (e.g., a single z-score or percentile-based index). This makes it difficult to distinguish the different ways novel contributions enter a field and influence how its knowledge structure develops. The ***second*** line of work advances multidimensional or typological frameworks. Foster and colleagues [2], for example, classify research strategies in biomedicine using Medical Subject Headings (MeSH), distinguishing conservative, mainstream, and risky strategies based on how frequently topic combinations have appeared in prior work. Here, risk is defined by the rarity of recombination within an existing topic space. Leahey and colleagues [17] differentiate novelty in theory, methods, and results by combining Citation Classics essays with citation data to examine how different types of novelty shape patterns of influence among highly cited papers across fields. While these studies move beyond scalar scores by recognizing that novelty is not homogeneous, their operationalizations either classify research according to the rarity within an existing topical space or examine novelty among already influential work. The ***third*** approach relies on disruptiveness-based metrics, which infer novelty from shifts in downstream citation patterns, whether a contribution amplifies or displaces prior work [18,19]. These measures capture important dynamics of knowledge development but remain inherently retrospective, as they depend on subsequent citation behavior [20]. More recently, high-dimensional semantic models have operationalized novelty as semantic distance or abnormality in textual space [16,21]; despite the methodological sophistication, these approaches also reduce novelty to a single quantity and are primarily optimized to predict citation impact rather than to characterize the diverse structural roles novelty may play as it emerges within a field.

Across these approaches, three primary limitations emerge. First, most metrics treat novelty as a unidimensional quantity, obscuring the possibility that different forms of novelty operate through distinct structural mechanisms. While a small number of studies introduce multidimensional distinctions, these dimensions are typically defined in terms of recombination frequency within an existing topic space or in term of substantive domains (e.g., theory, method, results) rather than in terms of how contributions structurally transform the evolving knowledge network. Second, many focus on high-impact or highly visible work, thereby overlooking early adopters, fast followers, or consolidating contributions that may be critical for field development [22,23]. Third, these measures are often weakly grounded in a general theory of novelty production, making it difficult to interpret what aspect of novelty a given metric actually captures [24]. These limitations are not simply technical. They reflect a broader mismatch between how novelty is produced, often through exploratory and uncertain processes, and how scientific evaluation is organized. By implicitly treating novelty as a precursor to success, these direct metrics still tend to privilege alignment with established knowledge structures over deviation from them, sidelining contributions that challenge dominant paradigms or institutional expectations [25]. This bias follows from a goal-oriented logic that weigh alignment with established paradigms more than ideas deviate from them. However, empirical evidence have shown that search processes that reward uniqueness without fixed objectives can outperform traditional, outcome-based strategies in certain complex problem-solving tasks [26].

In response to these limitations, this paper introduces an alternative, theory-driven typology of novelty grounded in the philosophy and sociology of science. We conceptualize novelty as a multidimensional process and develop a triadic framework that distinguishes among different structural pathways through which new ideas emerge, are reconfigured, and become stabilized within a field. Rather than inferring novelty from later success, the framework measures novelty at the point of contribution and traces its subsequent embedding in the evolving knowledge structure. The paper proceeds as follows. We first articulate the conceptual dimensions that motivate the triadic framework and derive the three novelty pathways. We then introduce a network-based operationalization that translates these concepts into measurable indicators of novelty. Next, we apply the framework to an interdisciplinary research field to examine how different novelty pathways relate to citation recognition. We extend the analysis by introducing a simulated baseline model that enables standardized comparison and aggregation beyond individual articles. Finally, we situate the framework relative to existing novelty measures and discuss implications for research evaluation and science and innovation policy.

## Conceptual Dimensions of Scientific Novelty

To move beyond scalar definitions of novelty, we identify four conceptual dimensions that characterize how novelty arises and is expressed in scientific research. These dimensions -- relational, multifaceted, contextual, and dependence on the first subsequent connections -- define novelty as a process embedded in knowledge structures over time, rather than as an intrinsic attribute of individual publications. Taken together, they follow a coherent generative logic: the relational dimension specifies what is novelty made of; the multifaceted dimension explains why not all relational novelties are the same; the contextual dimension explains why some novelties are recognized while others are ignored; and dependence on subsequent connections explains how novelty becomes embedded in the knowledge structure of a field.

**Relational**. Novelty does not arise from isolated ideas, but from new relationships among existing ones. Building on Bacon's view of scientific discovery, Merton [4] argued that novelty emerges from sustained interactions between scientists and existing knowledge, rather than from individual efforts alone. This relational view was further articulated by Swanson's [27] hypothesis of undiscovered public knowledge, which shows how significant discoveries often result from linking disconnected literatures. For examples, in bioinformatics, the integration of biological data with computational methods transformed genetic research, as exemplified by the Human Genome Project [28]. Similarly. in econophysics, tools from physics have been applied to model complex economic systems, offering new insights into market behavior [29]. These cases illustrate that novelty is generated through relational reconfiguration within knowledge systems, not through isolated intellectual acts.

**Multifaceted**. All relational novelties are not created equal. They take multiple forms, shaped by the varied roles that scientists play in generating, organizing, and advancing knowledge. Drawing on Merton's [4] discussion of scientific labor and Znaniecki's [30] typology of intellectual roles, novelty can emerge from initiating new lines of inquiry, synthesizing disparate knowledge, or deepening existing paradigms. Burt's [31] brokerage theory similarly highlights how bridging structural holes in knowledge networks leads to different types of informational advantages. These distinctions are evident in the varying logics of multidisciplinary, interdisciplinary, and transdisciplinary research, which represent additive, interactive, and integrative approaches to connecting knowledge across fields [32]. Thus, novelty cannot be captured by a single mechanism, rather, it varies depending on how ideas are combined and by whom.

**Contextual**. Novelty is inherently contextual, shaped by the specific conditions under which it emerges and is recognized. Merton [4] noted that multiple discoveries occur independently and simultaneously, suggesting that breakthroughs are often conditioned by the readiness of the knowledge system, much like ripe apples falling from a tree [14]. Truly singular discoveries are rare and require explanation beyond individual genius. The recognized novelty of an idea depends on factors such as timing, location, and disciplinary context. For example, artificial intelligence (AI) research has existed for decades, but only gained widespread recognition after advances in computing hardware (e.g., GPUs) and algorithmic design [33]. Even today, innovations in AI may be seen as novel in some domains or regions while appearing routine in others [34,35]. These examples highlight that novelty cannot be separated from the contextual conditions that shape its recognition and uptake.

**Dependence on First Subsequent Connections**. Novelty is rarely stabilized through isolated acts; rather, it depends on first subsequent connections that reinforce, extend, or embed the initial contribution within the knowledge structure. Merton's [4] concept of "multiples" points to the redundancy in scientific discovery, which is often viewed as wasteful duplication. However, Burt [31] argues that such redundancy can play a generative role by stabilizing emerging ideas and filling structural gaps. In many cases, a novel contribution may not become cognitively integrated or durable until it is supported by additional follow-up efforts. For example, Einstein's theory of gravitational waves was proposed in 1916 but remained speculative until confirmed by the LIGO and Virgo collaborations a century later [36]. Similarly, the Higgs boson hypothesis gained prominence only after empirical validation at the Large Hadron Collider [37]. These cases show that first reinforcement efforts are not indicators of success, but a structural condition through which novelty becomes embedded in the field knowledge structure.

Together, these dimensions describe novelty as a process that originates in new relationships among ideas (relational), manifests in multiple structural forms (multifaceted), is recognized unevenly across contexts (contextual), and becomes legible and durable only through first subsequent connections.

## A Triad of Novelty Typology: Pioneer, Maverick, and Vanguard

These four conceptual dimensions jointly imply that novelty cannot be expressed in a single form. If novelty arises through relational reconfiguration, manifests in multiple structural roles, is recognized contextually, and stabilizes through subsequent connections, then it must enter the knowledge system through distinct pathways. These pathways reflect different ways in which research contributions alter the structure of a knowledge field.

At a structural level, novelty can affect a knowledge system in only three fundamental ways: by introducing new elements, by creating new relationships among existing elements, or by reinforcing previously weak relationships. These three modes correspond directly to the triadic typology developed here. We refer to these pathways as Pioneer, Maverick, and Vanguard novelty.

*Pioneer novelty* captures contributions that expand the conceptual space of a field by introducing new topics. These works add previously absent elements to the knowledge structure, opening new areas of inquiry and redefining the boundaries of a field. Pioneer novelty reflects the generative role of initialing new domains, but such contributions often face substantial legitimacy barriers at the time of their introduction.

*Maverick novelty* captures contributions that reconfigure existing knowledge by linking previously distant or disconnected topics. Rather than expanding the topic space, Mavericks reshape the internal structure of the field by creating novel pathways between established areas. This form of novelty reflects recombination across boundaries and often challenges dominant organizational or disciplinary silos.

*Vanguard novelty* captures contributions that stabilize emerging ideas by reinforcing weak but existing connections. These works do not introduce new topics or unprecedented links; instead, they consolidate fragile relationships that have appeared only sporadically. Vanguard novelty reflects the role of early adopters and fast followers in making new ideas cognitively integrated and durable within a field.

Together, these three pathways exhaust the structural possibilities through which novelty can enter and reshape a knowledge system. Pioneer novelty introduces new elements, Maverick novelty reconfigures existing relationships, and vanguard novelty stabilizes the new structure. Importantly, these forms of novelty are analytically distinct from any eventual impact on popularity. Each pathway reflects a distinct mechanism and has context-dependent impact on eventual recognition.

## Operationalizing the Triadic Novelty Framework

To operationalize the triadic novelty framework, we represent the evolving structure of a field as a weighted co-occurrence citation network (WCCN). In this network, nodes correspond to research topics, and weighted edges represent the frequency with which pairs of topics co-occur within individual publications. This representation captures the *relational* dimension of novelty by embedding each contribution within the existing knowledge structure at the time of publication.

For each novelty type, we develop two complementary measures. The *initial novelty score* captures the immediate structural contribution an article makes to the knowledge network. The *novelty impact score* captures whether and to what extent that initial contribution becomes embedded in the evolving structure of the field over a subsequent time window. Importantly, novelty impact is a descriptive measure rather than a causal one. It reflects the persistence and diffusion of a structural contribution under the influence of multiple exogenous factors (e.g., institutional dynamics, funding priorities, and the broader field constitutes exogenous factors). These operationalizations are illustrated in Figure 1.

### *Pioneer Novelty: Introducing New Topics*

**Pioneer novelty** captures contributions that introduce entirely new topics into the knowledge structure of a field. These pioneering articles expand the boundaries of scientific inquiry by adding new conceptual areas, represented as previously unseen nodes in the topic network. In Figure 1: Panel I, node F (orange) illustrates a new topic introduced into an existing network (depicted by blue nodes and gray ties).

**Initial Pioneer Novelty Score.** Let $I_r$ denote the set of topics associated with the article $r$, and let $R_t$ be the set of all articles published up to time $t$. We define Pioneer novelty $N_{Pioneer}$(r) as the number of new nodes introduced by the article $r$. That is, topics represented in $I_r$ but absent from the set $I_{R_t}$ comprising all the topics introduced before the time it was published. Formally:

$$N_{Pioneer}(r) = |I_r \backslash I_{R_t}|$$

This score reflects the breadth of conceptual novelty contributed by the article, capturing the extent to which it expands the topical space of the field.

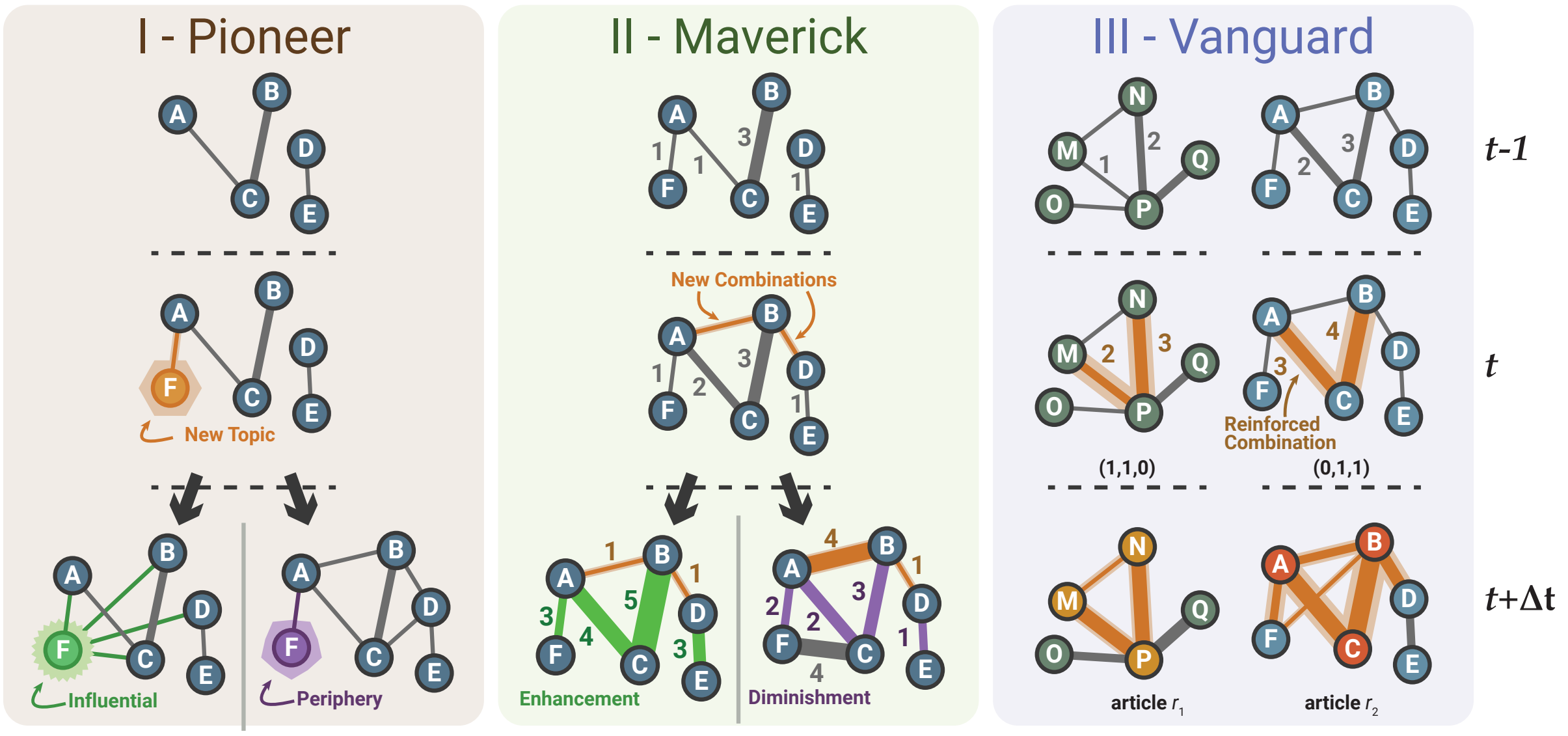

***Figure 1. Schematic Representation of the Triadic Novelty Typology.*** *This figure illustrates three types of research novelty – Pioneer, Maverick, and Vanguard – using snapshots of the topic co-occurrence network across three points in time* $(t-1, t, t+\Delta t)$*. Nodes represent topics, and ties represent co-citation relationships, with tie weights indicating the number of articles co-citing each topic-pair.* ***Panel I (Pioneer Novelty):*** *This panel shows the introduction of a new topic at the time* $t$ *(highlighted in orange node F). Based on its subsequent influence on the network, it could be an Influential (green node) or a Peripheral Pioneer (purple node).* ***Panel II (Maverick Novelty):*** *This panel shows the introduction of a new link (highlighted in orange) between distant nodes A and B at the time* $t$*, reducing their shortest path distance. This novelty may lead to either Enhancement (green ties) or Diminishment (purple ties) of existing ties.* ***Panel III (Vanguard Novelty):*** *This panel shows the reinforcement of existing weak ties (indicated by thickened orange lines). Its impact is quantified by the average change in the weighted degree centrality of the connected nodes before and after the reinforcement.*

**Pioneer Novelty Impact Score**. Not all new topics gain traction [38]. Some pioneering ideas become central to future research (e.g., green node F in Figure 1: Panel I), while others remain peripheral (e.g., purple node F). The Pioneer impact Score $S_{Pioneer}(r)$ captures the maximum[a] uptake of any newly introduced topic by article $r$ over a future window, indicating whether the expanded conceptual space becomes integrated into subsequent research. Specifically:

$$S_{Pioneer}(r) = \max_{i \in I_r} \frac{|R_{t,t+\Delta t}(i)|}{|R^{total}_{t,t+\Delta t}|}$$

Where $R_{t,t+\Delta t}(i)$ denotes the set of articles published within the future window $(t, t+\Delta t)$ that are associated with the topic $i$ , and $|R^{total}_{t,t+\Delta t}|$ represents the total number of articles published during that same period.

## *Maverick Novelty: Reconfiguring Existing Knowledge*

**Maverick novelty** captures contributions that challenge the status quo by connecting previously unrelated or indirectly linked topics. Rather than introducing entirely new areas, Mavericks reshape the internal structure of the field by forging conceptual bridges across topical boundaries. In Figure 1: Panel II, the orange ties between nodes A-B and B-D represent new ties that shorten the conceptual distance between those topic pairs.

[a] The Pioneer novelty impact score can be measured as the average, median, maximum, or minimum of these changes; we choose the maximum for this study.

**Initial Maverick Novelty Score**. This novel contribution is computed as the average fractional (i.e., normalized between 0 and 1) change in shortest-path distance (i.e., the minimum number of ties to bridge the gap) caused by the newly introduced ties. Let $J_r$ denote the set of new ties introduced by the article $r$, and let $d_j$ represent the initial shortest distance path between the two nodes connected by the tie $j$, computed using the Dijkstra algorithm [39]. If two nodes are disconnected, their distance is treated as infinite. The distance change for each new tie is calculated as $1-\frac{1}{d_j}$ where $\frac{1}{\infty}=0$. The initial Maverick novelty score for a paper $r$ is defined as:

$$N_{Maverick}(r)=\frac{1}{|J_r|}\sum_{j\in J_r}\left(1-\frac{1}{d_j}\right)$$

A higher score indicates that the article introduces more conceptually distant connections, reflecting greater structural redirection.

**Maverick Novelty Impact Score**. New ties can redistribute attention within the surrounding knowledge structure. Because tie weights generally increase as the literature expands, we measure impact in growth-adjusted terms. We therefore distinguish between two descriptive impact patterns: *enhancement*, in which neighboring ties gain more growth-adjusted weight, and *diminishment*, in which attention is redirected away from adjacent structures. These reflect changes[b] in the proportional (i.e., publication-normalized) weights of neighbor ties. That is, ties connected to the nodes involved in the new link, before and after the publication of the Maverick article. In Figure 1: Panel II, the new ties connect nodes A, B, and D. The relevant neighbor ties include A-F, A-C, B-C, and D-E. If the average proportional weight of these ties increases after the new tie is introduced, the article exhibits *enhancement*; if it decreases, it exhibits *diminishment.* Assuming that each new tie is introduced by a different new article and $\Delta t=1$, in this scenario, the proportional weight change from $t-1$ to $t$ of the four neighbor ties was ½ (the only weight change A-C divided by the two new articles). From $t$ to $t+\Delta t$ this increases to 1 in the enhancement scenario and decreases to $1/8$ in the diminishment scenario. Thus, the enhancement score is calculated as ⅛ $=1/[4(1-½)]$ and the diminishment score is $3/32=$ $(¼)(1-½)$.

Let $H_r$ denote the set of neighbor ties to the article $r$. Let $\Delta W_{h,t}$ denote the weight gained by the neighbor tie $h$ ($\in H_r$) between time $t$ and $t+1$. Let $|R_t|$ denote the total number of articles at time $t$. Let $g_h^{before}=\frac{1}{\Delta t}\sum_{\tau=t-\Delta t}^{t-1}\frac{\Delta W_{h,\tau}}{|R_\tau|}$ and $g_h^{after}=\frac{1}{\Delta t}\sum_{\tau=t+1}^{t+\Delta t}\frac{\Delta W_{h,\tau}}{|R_\tau|}$ be the proportional weights of neighbor ties $h$ before and after the new ties. Let $\Delta g_h=g_h^{after}-g_h^{before}$. Enhancement occurs when $\Delta g_h>0$, while diminishment occurs when $\Delta g_h<0$. Hence, the function describing the measure is as follows:

[b] The Maverick novelty impact score can be measured as the average, median, maximum, or minimum of these changes; we choose the average for this study.

$$S_{Enhancement\ Maverick}(r) = \frac{1}{|H_r|} \sum_{h \in H_r} max(\Delta g_h, 0)$$

$$S_{Diminishment\ Maverick}(r) = \frac{1}{|H_r|} \sum_{h \in H_r} min(\Delta g_h, 0)$$

The enhancement score reflects whether a Maverick connection increases engagement with existing related ties, while the diminishment score captures a decline in attention to previously adjacent topics.

### *Vanguard Novelty: Reinforcing Emerging Ideas*

**Vanguard novelty** captures contributions made by early adopters or fast followers that reinforce weak but existing connections between topics. Because one article may strengthen multiple ties, Vanguard novelty is evaluated relative to the prior frequency of each reinforced connection, with higher novelty assigned to articles that consolidate rarer ties. In Figure 1: Panel III, the connection between nodes M and P or N and P illustrates ties that had appeared only once before and are strengthened by a new article.

**Initial Vanguard Novelty Score**. The process involves comparing the prior frequency of each strengthened tie and rank articles accordingly, up to a predefined limit $K$, beyond which ties are no longer considered novel. Let $U_r = (u^1, \ldots, u^k, \ldots, u^K)$ denote the ranking vector for the article $r$, where $u^1$ is the number of ties strengthened by $r$ that appeared once before time $t$, $u^2$ the number that appeared twice, and so on. The ranking vector allows for lexicographic ordering [40] of Vanguard articles. If two articles differ in $u^1$, the one with the higher value is ranked higher. If they have the same $u^1$, we compare $u^2$, and so on. Articles with identical ranking vectors have the same Vanguard novelty. This lexicographic process is similar to alphabetizing words; the first "letter" ($u^1$) takes priority, and subsequent "letters" ($u^2$, $u^3$, etc.) are considered only if needed. Once articles are ranked lexicographical, ordinal numbers are assigned with "1" given to the top-ranked article(s), "2" to the next, and so on. For example, in Figure 1: Panel III, the hypothetical article $r_1$ has ranking vector (1,1,0). Another article $r_2$, within strengthens two ties that previously appeared twice (i.e., A-C) and three times (i.e., B-C) has ranking vector (0,1,1). If two articles strengthen links appearing once and twice before in identical counts, we then compare whether one also strengthens a tie that appeared three times. With $K$ = 5, ties appearing more than five times are excluded from consideration; under this rule, $r_1$ranks higher than $r_2$.

**Vanguard Novelty Impact Score**. Because Vanguard ties increase node connectivity, they could also increase the influence of the reinforced links around them. Its significance is thus measured by the average[c] change in the weighted degree centrality of the nodes involved. Let $E_r$ represent the set of topics in the article $r$ published at the time $t$ connected by Vanguard ties. Let $w_{e,t}$ denote the total count of articles citing a topic $e$ ($\in E_r$) and $|R_t|$ the total number of articles at the time $t$. The Vanguard Novelty Impact Score is:

[c] The Vanguard novelty impact score can be measured as the average, median, maximum, or minimum of these changes; we choose the average for this study.

$$S_{Vanguard}(r) = \frac{1}{|E_r|}\sum_{e \in E_r}\left(\frac{w_{e,t+\Delta t}}{|R_{t+\Delta t}|} - \frac{w_{e,t-1}}{|R_{t-1}|}\right)$$

This measure reflects whether the connections reinforced by Vanguards become more structurally central over time, capturing their increasing integration into the network of the field.

## Research Context: Philanthropic and Nonprofit Studies as a Testing Field

We apply the triadic novelty framework to philanthropic and nonprofit studies (PNPS), an interdisciplinary and evolving research domain that spans the social sciences, life sciences and biomedicine, technology, arts and humanities, and physical sciences [41]. PNPS is characterized by intellectual diversity, evolving disciplinary boundaries, and heterogeneous publication outlets, making it a theoretically informative context for examining how different forms of novelty arise and are recognized [41].

Importantly, PNPS functions as a testing field rather than an idiosyncratic case. Compared with more highly institutionalized disciplines (i.e., Physics, Economics), PNPS exhibits less consolidated evaluation norms and weaker legitimacy boundaries [41,42]. This structural openness allows novel contributions to compete for attention before recognition mechanisms are fully stabilized, making it easier to observe how different novelty pathways unfold [3]. If the triadic framework were unable to distinguish Pioneer, Maverick, and Vanguard novelty in such a setting, it would be unlikely to do so in more rigid fields. Conversely, evidence of systematic differences among novelty pathways in PNPS would be consistent with the framework capturing general structural mechanisms rather than field-specific artifacts.

PNPS is also substantively relevant to broader concerns of science, technology, and innovation (STI) policy [43]. Research in this domain frequently engages with social problems, organizational design, and public value [41,44], areas in which judgments about novelty, legitimacy, and impact are especially consequential. As a result, insights from this context speak not only to the development of PNPS as a field, but also to how emerging and interdisciplinary areas of research are evaluated, supported, and rewarded more broadly.

## Results

We examined the triadic novelty framework in a dataset of 41,623 peer-reviewed articles published between 1960 and 2017 in the PNPS field. These articles cite 313,172 unique references indexed in the Web of Science (WoS). The dataset was drawn from a broader corpus of 60,684 articles spanning 1899 - 2022, with pre-1960 publications used to construct the knowledge base of the field. Because we assessed citation impact using a five-year forward citation window, the regression analysis included only articles published no later than 2017. Citation counts reflect citations from other WoS-indexed publications.

Our results are organized into four parts. We first assessed the alignment between initial novelty and novelty impact across novelty pathways. We then examined how these pathways are reflected in citation counts, treating citations as an institutional signal of recognition. Next, we introduced a simulated baseline model to account for structural differences across time and fields, enabling standardized comparison and demonstrating how the framework can be applied beyond individual articles. Finally, we validated the triadic novelty measures with an established novelty proxy to clarify how the framework relates to existing scalar approaches.

### *Alignment between Initial Novelty and Novelty Impact Scores*

We examined how initial novelty scores align with corresponding novelty impact scores, which capture the extent to which initial structural contribution becomes embedded in the knowledge structure of the field. Alignment between these measures indicates whether novelty at the point of entry translates into subsequent structural uptake, and whether this relationship differs across novelty types. To assess this alignment, we compute Spearman correlations using only non-zero values for each novelty type; for Maverick novelty, we use the absolute value of diminishment impact scores, as smaller raw diminishment values indicate greater structural impact.

As shown in Table 1, alignment between initial novelty and novelty impact varies across pathways. Mavericks exhibit the strongest associations, particularly for diminishment-based impact ($\rho = 0.29$), indicating that articles introducing distant conceptual connections are more likely to see those contributions integrated over time. Pioneers show only a weak association between initial novelty and subsequent embedding ($\rho = 0.087$), suggesting that introducing new topics does not reliably translate into broader structural uptake. Vanguards exhibit a modest negative relationship ($\rho = -0.131$), indicating that initial reinforcement of weak ties does not necessarily translate into broader structural uptake.

***Table 1. Correlation Between Initial and Impact Novelty Scores by Type***

| Novelty Type | Spearman Correlation (ρ) |
|---|---|
| **Pioneer** | 0.087 |
| **Maverick (Enhancement)** | 0.216 |
| **Maverick (Diminishment)** | 0.290 |
| **Vanguard** | –0.131 |

### *Citation Impact of Novelty Types*

We next examine how different novelty pathways are recognized within prevailing scientific reward systems—citation impact. To do so, we estimate mixed-effects negative binomial models predicting five-year citation counts as a function of novelty type and corresponding novelty impact scores, controlling for publication year, journal, and field. Descriptive statistics for regression variables are reported in Figure S1, and full model results are presented in Table 2.

**Pioneer Novelty: New Topics Often Went Unrewarded.** Pioneer novelty, which captures the introduction of new topics into the field network, showed no significant relationship with five-year citation counts. As shown in Table 2, neither the direct effect of introducing new topics (regression coefficient $\beta = -0.014, p - value = 0.84$) nor the interaction effect with Pioneer novelty impact ($\beta = 0.104, p = 0.17$) was significant. However, the predicted citation (Figure 2A) and marginal effect plots (Figure 2B) revealed a more nuanced picture. Figure 2A shows that pioneer novelty boosted citations only when paired with high levels of novelty impact. When the uptake is low, novelty has little effect or even reduces citations. Figure 2B further shows that the marginal effect of pioneer novelty stayed near zero at low impact levels but became more positive as impact increased. This pattern indicates that frontier-expanding contributions tend to receive citation recognition only when others subsequently adopt and build upon the newly introduced topics.

***Table 2. Mixed Effect Negative Binominal Regression Results.***

| | DV = 5-Year Citation Counts | | |
|---|---|---|---|
| **Fixed Effect Predictors** | **Coefficient** | **SE** | **95%CI** |
| Pioneer Novelty | -0.014 | 0.071 | [-0.152, 0.124] |
| Pioneer × Pioneer Impact | 0.104 | 0.075 | [-0.043, 0.250] |
| Maverick Novelty | 0.546*** | 0.057 | [0.435, 0.658] |
| Maverick × Enhancement | 0.053 | 0.199 | [-0.337, 0.442] |
| Maverick × Diminishment | 0.279 | 0.383 | [-0.472, 1.030] |
| Vanguard Novelty | 0.109*** | 0.003 | [0.103, 0.115] |
| Vanguard × Impact | 0.026*** | 0.002 | [0.021, 0.030] |
| Vanguard × Impact^2 | -0.005*** | 0.001 | [-0.007, -0.003] |
| Field: Social Sciences | 0.410*** | 0.051 | [0.310, 0.510] |
| Field: Arts & Humanities | -0.780*** | 0.064 | [-0.905, -0.655] |
| Field: Life Sci. & BioMed | 0.819*** | 0.051 | [0.719, 0.918] |
| Field: Technology | 0.541*** | 0.068 | [0.408, 0.675] |
| **Random Effect Predictors** | **Variance** | **SD** | **95%CI** |
| Year | 0.382 | 0.618 | [0.506, 0.755] |
| Journal | 1.035 | 1.017 | [0.989, 1.046] |

*N = 40,707*
** p < 0.05, ** p < 0.01, *** p < 0.001*

**Maverick Novelty: Recombination Was Rewarded, Especially When It Redirected.** Maverick novelty captures research that forged new ties between previously disconnected or distantly linked topics. The direct effect was strongly positive ($\beta = 0.546, p < 0.001$) indicating that these novel connections tended to be well-received. Although its interaction with enhancement ($\beta = 0.053, p = 0.79$) and diminishment ($\beta = 0.279, p = 0.47$) impact scores were not significant, the predicted citation plots (Figures 2C and 2E) revealed that higher Maverick novelty scores consistently led to more citations across levels of both enhancement and diminishment impact. The marginal effect plots (Figures 2D and 2F) further clarify this pattern. For enhancement impact (Figure 2D), the marginal effect of Maverick novelty remained relatively stable and largely positive, suggesting that recombination is reliably recognized even when it does not reinforce surrounding structure. By contrast, the marginal effect increases with higher diminishment impact (Figure 2F), indicating that Maverick novelty tended to receive greater recognition when it redirects attention away from existing topic structure.

**Vanguard Novelty: Reinforcement Paid off, but Only to a Point.** Vanguard novelty captures research that strengthens weak existing topic connections. The predefined limit $K$ was set as 5. Vanguard novelty is positively associated with citation ($\beta = 0.109, p < 0.001$), and this association strengthens with higher novelty ($\beta = 0.026, p < 0.001$). As depicted in Figure 2G, predicted citations increase as reinforced topic pairs become structurally central in the field.

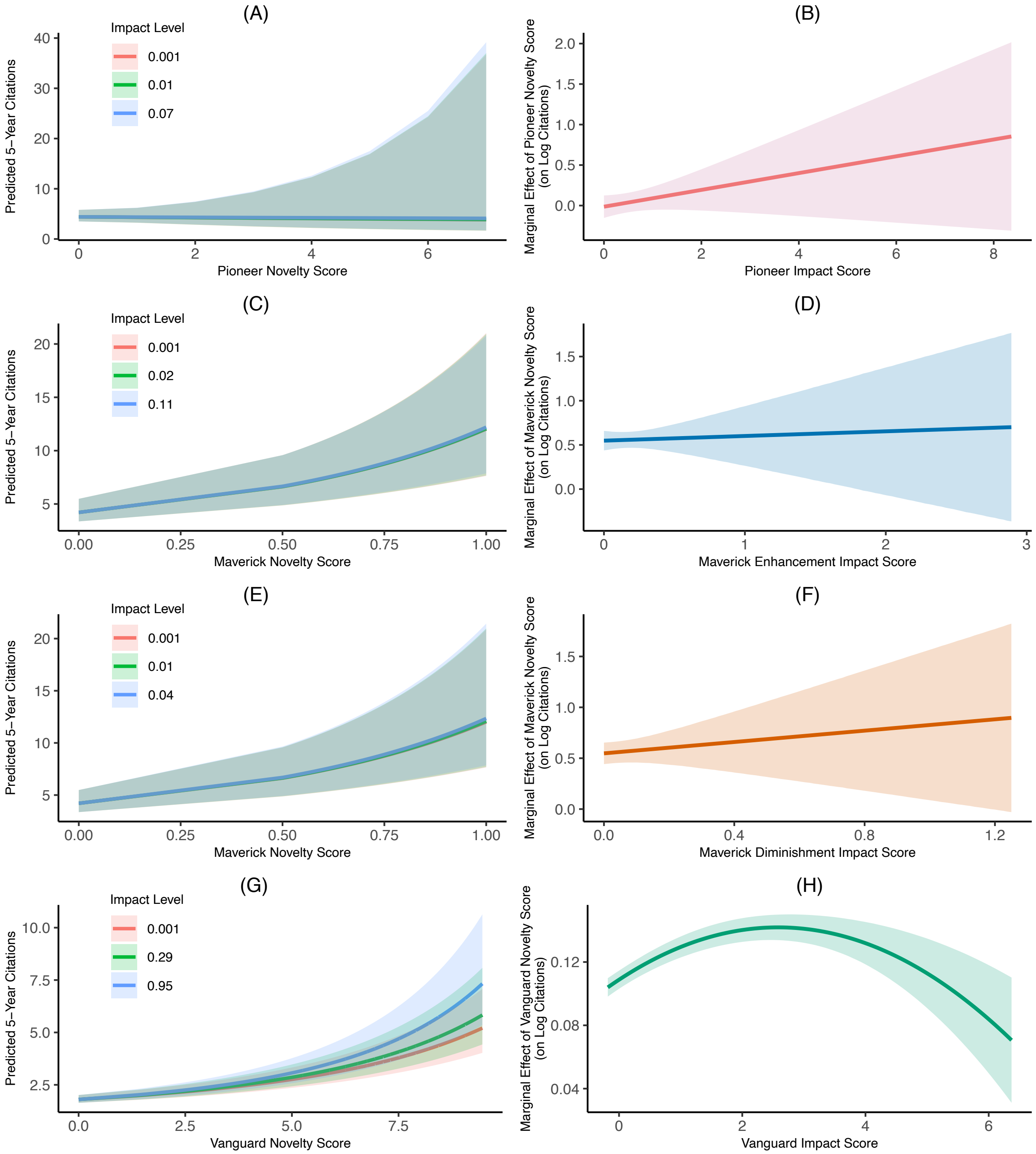

***Figure 2. Predicted Citations and Marginal Effects by Novelty Type and Impact Interaction.*** *Panels A, C, E, and G show predicted five-year citation counts across levels of Pioneer, Maverick (Enhancement and Diminishment), and Vanguard novelty. Impact values are set to representative low (0.001), average (mean), and high (mean +1SD) levels based on the empirical distribution. Panels B, D, F, and H display the marginal effects of novelty scores across the full range of corresponding impact variables. Shaded bands represent 95% confidence intervals.*

However, this effect exhibits diminishing returns. The significant negative quadratic term ($\beta = -0.005, p < 0.001$) indicates that the citation benefit of Vanguards peaks at moderate levels of novelty impact and declines thereafter. Figure 2H illustrates this hill-shaped pattern:

reinforcement is most strongly recognized when it helps stabilize emerging connections, but additional reinforcement in already well-integrated areas yields less citation recognition.

In summary, these results confirm the theoretical claim of the framework: novelty is not a single trait, but a multidimensional process shaped by how ideas enter, restructure, and become embedded within a field. Consistent with this view, different novelty pathways exhibit distinct patterns of recognition. Maverick novelty is consistently rewarded, especially when it redirects attention within the knowledge network. Vanguard novelty is recognized when it consolidates emerging structures, but its benefits diminish as those structures become established. In contrast, Pioneer novelty often goes under-recognized unless newly introduced topics are subsequently taken up and integrated into the field. Together, these patterns underscore that recognition systems do not respond uniformly to novelty but selectively reward certain pathways of knowledge change over others.

### *Simulated Baseline Model and to Journal-Level Novelty*

Because both citation volume and novelty opportunities vary systematically over time and across fields, raw novelty scores are not directly comparable across articles. Later publications face a denser knowledge landscape, and fields differ in size and topical breadth. To account for these structural differences, we constructed a simulated baseline model that allows novelty levels and novelty impact to be compared on a common scale.

The baseline model was adjusted for citation inflation [45], preferential attachment [46–48], and the limited number of WoS subject categories. For each publication year, the model generated the same number of articles as in the observed data. Each simulated paper was assigned the same number of subject categories as its real counterpart, ensuring structural consistency. Like Price's model of cumulative advantage [48], the selection of subject categories followed a preferential attachment process. The resulting distributions of novelty scores provide a reference against which observed article-level novelty is standardized using z-scores.

We then used these standardized novelty scores to illustrate how the triadic framework can be applied beyond individual articles, using journals as an example of group-level assessment. Aggregating standardized novelty scores across articles, we identified which journals tend to publish highly novel work. We ranked the journals that contain at least 30 PNPS articles across three types of novelty based on initial novelty scores. Table 3 reports the top ten journals for each novelty type based on average standardized novelty scores. These ranking include both interdisciplinary and field-specific outlets, demonstrating that high novelty is not confined to traditional PNPS journals. Notably, several journals (e.g., Public Relations Review, Social Science & Medicine, etc.) consistently appear across different novelty types, showing cross-field capacity to support novel efforts.

To further illustrate these patterns, Figure 3 visualizes the average standardized novelty and novelty impact scores for the three core journal in PNPS [41]. *VOLUNTAS* exhibits the highest Pioneer novelty, consistent with its comparative and international scope, but relatively low Pioneer impact, indicating limited subsequent uptake of newly introduced topics. In contrast, *Nonprofit and Voluntary Sector Quarterly (NVSQ)* ranked lower on Pioneer novelty but achieved the highest Pioneer impact, suggesting greater selectivity in frontier-expanding contributions coupled with stronger structural embedding. These contrasting profiles illustrate how journals may play different roles in field development, either by broad exploration of new topics or by selectively amplifying those that gain traction.

*Table 3. Journal Rankings by Standardized Novelty in Philanthropic and Nonprofit Studies.*

| | Pioneer | Maverick | Vanguard |
|---|---|---|---|
| **1** | PUBLIC RELATIONS REVIEW | SOCIAL SCIENCE & MEDICINE | AMERICAN NATURALIST |
| **2** | JOURNAL OF ECONOMIC PSYCHOLOGY | REVESCO-REVISTA DE ESTUDIOS COOPERATIVOS | INTERNATIONAL JOURNAL OF BEHAVIORAL DEVELOPMENT |
| **3** | SOCIETY & NATURAL RESOURCES | SOCIETY & NATURAL RESOURCES | EVOLUTION |
| **4** | SOCIAL DEVELOPMENT | SOCIAL WORK | FORBES |
| **5** | PHYSICA A-STATISTICAL MECHANICS AND ITS APPLICATIONS | ZYGON | HARVARD BUSINESS REVIEW |
| **6** | BEHAVIORAL ECOLOGY | COMPUTERS IN HUMAN BEHAVIOR | ANIMAL BEHAVIOUR |
| **7** | AMERICAN SOCIOLOGICAL REVIEW | INTERNATIONAL JOURNAL OF SOCIAL ECONOMICS | JOURNAL OF TAXATION |
| **8** | SOCIAL SCIENCE & MEDICINE | SOCIAL MOVEMENT STUDIES | BEHAVIORAL ECOLOGY AND SOCIOBIOLOGY |
| **9** | BEHAVIORAL ECOLOGY AND SOCIOBIOLOGY | PUBLIC RELATIONS REVIEW | BEHAVIORAL ECOLOGY |
| **10** | WORLD DEVELOPMENT | COMMUNITY DEVELOPMENT | JOURNAL OF EVOLUTIONARY BIOLOGY |

## *Comparing with the Disruption Index*

As a robustness check, we compared our standardized and raw novelty scores to a widely used novelty proxy, the disruption index [18,19]. Table 4 shows the Spearman correlation between our novelty metrics and normalized disruption scores (log-transformed).

Overall, the correlations indicate limited overlap between the two approaches. For standardized scores, correlations are close to zero across all novelty types (e.g., $\rho = -0.041$ for Pioneer novelty), suggesting that the triadic framework captures dimensions of novelty not reflected in the disruption index. Correlations are somewhat stronger when raw scores are used, particularly for Maverick novelty ($\rho = 0.399$), indicating that the disruption index may partially reflect structural recombination among existing topics. By contrast, correlations remain weak for Pioneer and Vanguard novelty, consistent with the view that the disruption index does not capture novelty associated with expanding the conceptual space of a field or reinforcing emerging connections.

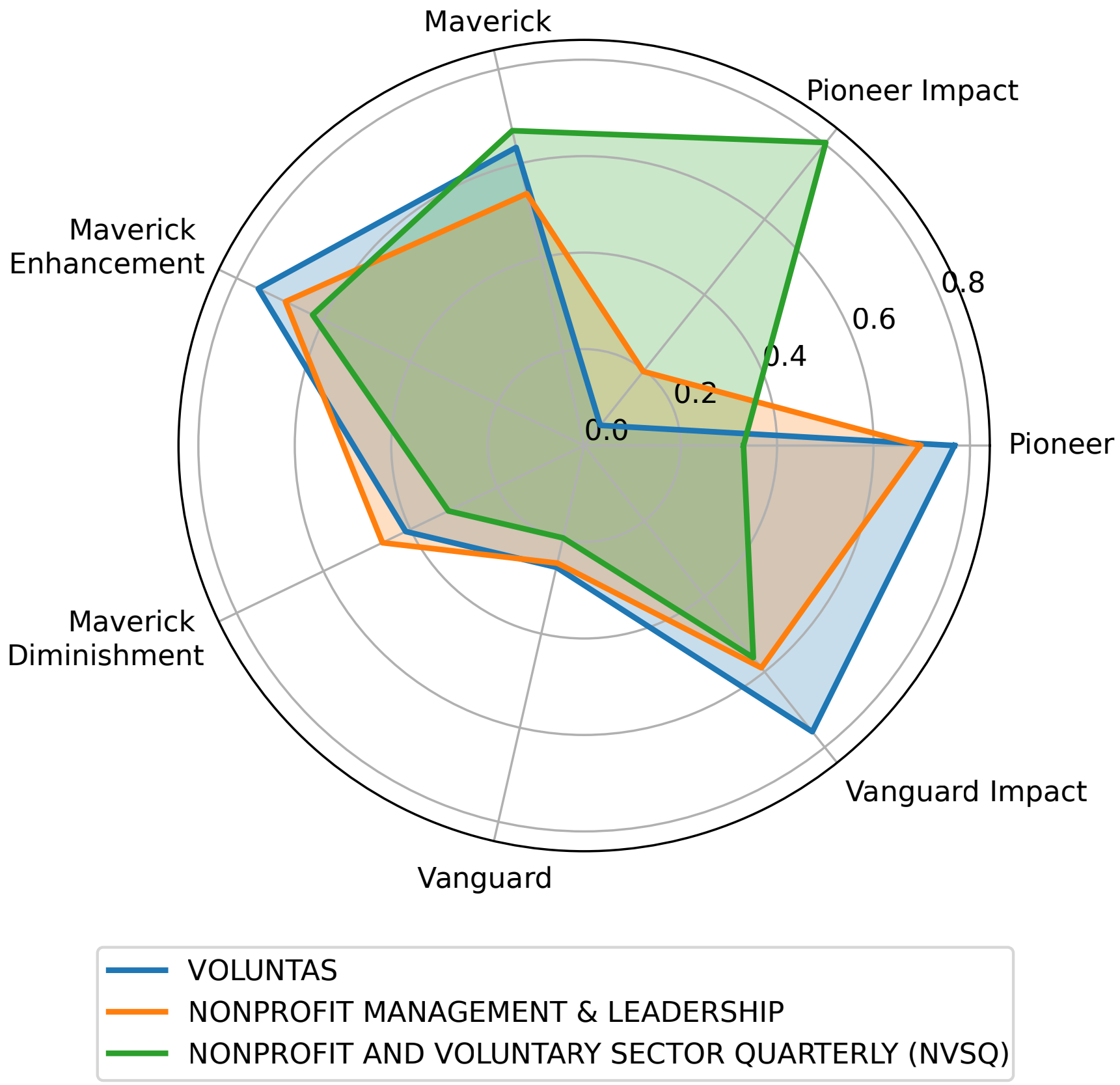

***Figure 3. Novelty Performances Across Core Journals in Philanthropic and Nonprofit Studies.*** *Each polar axis corresponds to a standardized novelty score or respective impact.*

These differences are expected given the distinct construction of the two measures. Disruption scores are not standardized against a simulated baseline and are known to be sensitive to citation inflation [20], which can bias comparisons across time. In contrast, our standardized novelty scores explicitly adjust for publication year and reference structure, which likely contributes to the lower observed correlations. Together, these patterns indicate that the triadic novelty framework and the disruption index respond to different structural features of scientific change, rather than providing alternative measurements of the same underlying construct.

***Table 4. Correlation with Disruption Index.***

| | **Standardized Scores** | **Raw Scores** |
|---:|:---:|:---:|
| **Pioneer** | -0.041 | -0.244 |
| **Pioneer Impact** | 0.323 | -0.055 |
| **Maverick** | -0.036 | 0.399 |
| **Maverick Enhancement** | -0.031 | 0.159 |
| **Maverick Diminishment** | -0.008 | 0.242 |
| **Vanguard** | 0.034 | 0.043 |
| **Vanguard Impact** | -0.198 | -0.127 |

**Note:** Only strictly positive values of novelty and disruption scores are included in the analysis (strictly negative for Maverick Diminishment).

## Discussion

**A Triadic Novelty Framework for Understanding Scientific Novelty.** This study advances a theory-driven framework for understanding scientific novelty as a multidimensional process rather than as a single latent quantity revealed through later success. We, conceptually and empirically, distinguished three structurally distinct pathways through which novel ideas enter and reshape a field: Pioneer novelty, which introduces new topics; Maverick novelty, which creates new connections between disconnected topics; and Vanguard novelty, which reinforces weak but emerging ties. These pathways reflect different intellectual strategies and face different conditions of recognition. For each novelty type, we also developed measures of novelty impact that trace how initial structural contributions are subsequently embedded within the knowledge network. Pioneer impact captures the uptake of newly introduced topics; Maverick enhancement and diminishment capture whether new combinations reinforce or redirect attention within the surrounding structure; and Vanguard impact captures increasing centralization of reinforced topic pairs. All measures were computed from the same WCCN, ensuring conceptual and methodological consistency across novelty types.

Importantly, the framework is not designed to replace existing novelty metrics or to reduce novelty into a single score. Instead, it provides a complementary lens that distinguishes among different structural forms of novelty and traces how they evolve within a field.

**Novelty and Popularity/Recognition Trap.** A central motivation of this work is to distinguish between novelty from popularity/recognition. Our results show that the relationship between novelty and popularity is conditional rather than uniform and depends on how novelty is structurally integrated into the existing knowledge system. The stronger alignment for Maverick novelty, particularly when new connections redirect attention within the knowledge network, suggest that recombinatory novelty is more readily recognized when it reshapes familiar structures. By contrast, the limited citation benefits of Pioneer novelty reflect persistence challenges for contributions that lie outside established frameworks. Vanguard novelty (fast following and early adoption) is more predictable in uptake, but once a contribution becomes common, its marginal value diminishes. Together, these findings indicate that evaluation systems relying on citation-based metrics tend to favor novelty that reconfigures or consolidates existing knowledge over novelty that expands the conceptual frontier.

**From Individual Contributions to Institutional Patterns.** The journal-level analysis illustrates how article-level novelty measures can be aggregated to examine institutional patterns of knowledge production. Beyond individual articles, the framework enables analysis of novelty at higher levels of aggregation. By introducing a simulated baseline model, we standardize novelty scores across time and reference structures, making it possible to compare aggregated novelty profiles without conflating novelty with publication volume, field growth, or historical timing. Rather than evaluating journals normatively, this approach shows how different venues contribute to field development through distinct forms of novelty, revealing patterns that are not visible through citation-based indicators. More broadly, this aggregation strategy provides a foundation for studying novelty portfolios across a range of organizational contexts, including journals, departments, funding programs, or research teams. By shifting the unit of analysis from individual contributions to institutional actors, the framework makes it possible to examine how organizational structures shape the pathways through which different forms of novelty are produced and sustained.

**Relation to Existing Novelty Metrics and Future Validation.** The comparison with the disruption index serves primarily to situate the triadic framework within the broader landscape of novelty measures. The limited overlap between the two approaches underscores that they are grounded in different conceptualizations and structural features of scientific innovation, highlighting the need for validation strategies that go beyond pairwise metric comparisons.

Future research can pursue several complementary approaches to validate the framework. First, *construct validation*[d] could be conducted through expert-based assessments (i.e., Guttman-type scaling [49]), for example by asking domain experts to evaluate whether publications classified into different novelty pathways exhibit the theoretically expected features of those pathways. Second, *internal validation*[e] could be conducted by examining the robustness of the typology to alternative network constructions, time window, or topic representations. Comparing novelty rankings generated under different parameterizations would help clarify the sensitivity of the measures to modeling choices while remaining within the same framework. *External validation*[f] could be conducted through applying the framework to additional fields with different epistemic cultures or topic granularities, allowing assessment of whether the same novelty pathways emerge and operate similarly elsewhere [10]. Other benchmarks, such as grant or award decisions, or downstream technological applications, could also be used to examine how different novelty pathways align with forms of recognition beyond citations. Overall, these validation strategies consider that novelty takes multiple forms and may operate differently across contexts. Rather than relying on a single benchmark, the triadic framework is evaluated by whether each novelty pathway behaves in ways consistent with its underlying theoretical definition, that is, how novelty enters, persists, and is recognized within scientific fields.

**Limitations and Suggestions for Future Research.** Reliance on WoS data imposes several limitations related to data availability and representation. WoS omits citations from books, policy reports, and many interdisciplinary outlets [50], potentially underrepresenting the impact of certain novel contributions, particularly in the social sciences and humanities. In addition, using WoS subject categories as proxies for topics introduces a constraint on granularity. With only 254 predefined categories, the system may fail to capture fine-grained thematic distinctions especially when conceptual novelty occurs at the subfield or problem level rather than at the level of broad disciplinary categories. The limited variance in Pioneer novelty scores may partially explain the weak association with short-term citation impact.

These limitations are tied to data availability rather than to the framework itself. Coarse-grained topic systems may be sufficient for detecting novelty in emerging or loosely structured fields like PNPS, but finer-grained representations are likely required in more established domains with entrenched paradigms. Future research could apply the framework using alternative topic models derived from abstracts, full texts, or curated classifications available in other databases (i.e., ANZSRC [51] available via the Dimensions database), extending its applicability across fields.

[d] Construct validity assesses whether the framework captures the intended theoretical notion of novelty.
[e] Internal validity refers to how well the data supports the study results.
[f] External validity tests generalizability beyond the original empirical setting.

Temporal scope also matters. Although the five-year citation window captures early recognition for many articles, some novel ideas, particularly those associated with Pioneer novelty, may follow longer and less predictable trajectories. Longer windows may better capture long-term influence but reduce timeliness for evaluative purposes. Future work could examine how novelty pathways unfold across different time horizons and institutional contexts, such as generalist versus specialist journals or open versus closed publication models.

Finally, while the framework captures key structural and temporal dynamics of novelty, it does not fully reflect the cultural or interpretive dimensions embedded in topic relationships [52,53]. Moreover, recognition of novelty often occurs through channels not captured by bibliometric data, including policy uptake, technological application, or public discourse. Integrating alternative data sources and qualitative tracing of idea trajectories would further enrich understanding of how different forms of novelty are recognized and valued across contexts.

**Implications for Research Evaluation and STI policy.** The framework reinforces the need to decouple novelty from popularity in research assessment. Citations capture recognition, but not novelty itself [5,10,12]. By providing citation-independent, structurally grounded measures of novelty, the framework offers policymakers and institutions a way to identify under-recognized research trajectories and to support long-term scientific development more deliberately.

More broadly, the triadic novelty framework has implications for STI policy, particularly for how research is evaluated, funded, and governed. A central concern in STI policy scholarship is that evaluation systems do not merely assess research outcomes but actively shape research trajectories by privileging certain forms of innovation over others. Indicators that emphasize short-term visibility or early recognition can unintentionally discourage exploratory and high-uncertainty research that expands the conceptual frontier [54].

Our findings provide empirical evidence for this concern. Novelty that introduces new topics remain weakly recognized unless it is subsequently embedded in the knowledge structure, making such contributions especially vulnerable in evaluation frameworks that rely on early success signals [55]. By contrast, recombinatory and consolidating forms of novelty align more closely with existing knowledge structures and are therefore more likely to be rewarded [2,5,25]. These patterns help explain persistent policy concerns about the under-support of high-risk and frontier-expanding research. From a policy perspective, distinguishing among novelty pathways enables a shift from project-level risk avoidance toward portfolio-level management of uncertainty [5,7,55]. Rather than implicitly favoring a single mode of innovation, funding and evaluation systems can be designed to balance exploratory, recombinatory, and consolidating contributions, an approach that is particularly relevant for interdisciplinary and emerging fields, where novelty often precedes institutional recognition.

## Materials and Methods

**Analysis Dataset**. We applied the Triadic Novelty framework to the PNPS, an emerging interdisciplinary field that investigates topics such as altruism, charity, volunteerism, nonprofits, civil society, and social economy [44]. The full dataset comprised 60,684 articles published from 1899 to 2022, with 513,406 references retrieved from the WoS Core Collection. We selected WoS due to its comprehensive coverage and standardized classification system [56]. Only documents labeled as "articles", including journal papers and conference proceedings with cited references, were included [50]. The dataset was assembled using a curated list of journals and keywords relevant to the field [57,58], detailed in the Supplementary Information (SI). Given the developmental history of the field, articles published before 1960 were treated as the foundational knowledge base. Novelty in later publications was measured relative to this base using the field citation network. Our primary analysis focused on 41,623 articles published between 1960 and 2017, citing 313,172 unique references indexed in WoS. The five-year forward citation window was used to calculate novelty impact scores, enabling consistent measurement of early-stage scholarly recognition.

**Network Construction**. The WCCN was operationalized by the co-cited WoS subject categories [59]. For instance, if an article cites a reference to the subject category 'Public Economics' and another to 'Applied Physics,' nodes 'Public Economics' and 'Applied Physics' were connected in the WCCN of WoS subject categories. The weight of each tie was determined by the total number of articles that co-cited the two subject categories. Specifically, in our dataset, one reference could have up to nine different subject categories. When a reference was linked to multiple categories, the categories were weighted equally. Novelty scores then calculated within this network. Although the 254 subject categories span all disciplines and may only loosely align with topical nuances of individual references, they still meaningfully reflect the broader orientations and disciplinary approaches [60,61], thereby serving as a useful indicator of the topical evolution of the field.

**Statistical Analysis.** To examine how different types of novelty are recognized within current academic reward systems, we used a mixed-effects negative binomial regression model to estimate their association with five-year citation counts. Citation counts are used exclusively as an outcome variable capturing recognition and are not involved in the construction of any novelty measures. This modeling approach is appropriate given the nature of our outcome variable: citation counts are non-negative integers with a distribution characterized by overdispersion (i.e., variance exceeds the mean), which violates assumptions of standard linear and Poisson models [62]. The negative binomial model accounts for this overdispersion by introducing an additional dispersion parameter, resulting in more accurate standard errors and valid statistical inference.

Given the hierarchical structure of our data, with articles nested within journals ($j$) and publication years ($i$), we included random intercepts for both journal and year to capture unobserved heterogeneity [63]. The random effect (shown in Table 2) for journals exhibited substantial variance ($\sigma^2 = 1.035$) indicating that journal venue played a significant role in citation performance, likely reflecting differences in visibility, disciplinary culture, and audience reach [64]. The variance for publication year was smaller but still meaningful ($\sigma^2 = 0.382$), suggesting modest temporal shifts in citation behavior due to evolving research priorities or external conditions. Including these random effects improved the precision of fixed-effect estimates by isolating the influence of novelty from contextual variation.

The dependent variable was the citation counts received within five years of publication, as recorded in the WoS, limited to citations from other WoS-indexed publications. Independent variables included initial novelty scores and novelty impact scores for each of the three novelty types, along with their interaction terms. To control for disciplinary variation, we included dummy variables based on the WoS domains [61]. The estimation is summarized as follows:

$$
\begin{aligned}
5 - year\ Citation\ Count_{ij} \sim\ & \beta_0 + \beta_1 N_{ij}^{Pioneer} + \beta_2\,(N_{ij}^{Pioneer} \times S_{ij}^{Pioneer}) + \\
& \beta_3 N_{ij}^{Maverick} + \beta_4\,(N_{ij}^{Maverick} \times S_{ij}^{Enh\ Maverick}) + \\
& \beta_5\, N_{ij}^{Maverick} + \beta_6\,(N_{ij}^{Maverick} \times S_{ij}^{Dim\ Maverick}) + \\
& \beta_7\, N_{ij}^{Vanguard} + \beta_8\,(N_{ij}^{Vanguard} \times S_{ij}^{Vanguard}) + \beta_9\,(N_{ij}^{Vanguard} \times S_{ij}^{Vanguard^2}) \\
& \beta_{10} Field_{socsci} + \beta_{11} Field_{a\&h} + \beta_{12} Field_{biomed} + \beta_{13} Field_{tech} + \\
& (1 \mid Journal_j) + (1 \mid Year_i) + \varepsilon_{ij}
\end{aligned}
$$

To improve interpretability and stabilize the scale of the regression coefficients, several transformations are applied. The Pioneer novelty impact score was multiplied by 100. Both Maverick Enhancement and Diminishment were rescaled by multiplying their absolute values by 100. For Vanguard novelty, the initial novelty score required additional processing. Articles with a raw Vanguard score vectors equal to (0,0,0) were excluded. The remaining articles were ranked based on their score vectors, then the ranks were inverted so that higher-ranked articles corresponded to higher numerical values, and then the values were log-transformed to standardize the distribution. The Vanguard impact score was also multiplied by 100.

**Baseline Model Design.** To enable meaningful comparisons of novelty scores across time and domains, we simulated a baseline model that captures expected levels of novelty under random assignment. This model generated synthetic papers for each year between 1960 and 2017 that mirrored the observed distribution in the real data. Each synthetic paper retained the same number of subject categories as its real counterpart. These subject categories are assigned stochastically, based on a preferential attachment mechanism: the probability of selecting a subject category $i$ is proportional to its cumulative prior citations, formally defined as node strength $S(i)$ from the previous year. To ensure that rarely cited categories still have a non-zero selection probability, we introduced a linear attractiveness term $A$. The resulting selection probability is modeled as $P(i) \sim S(i) + A$.

We tested several values of $A$ and selected the one that best replicated the temporal trend of novelty scores in the real data. For each value of $A$, we ran 100 iterations of the baseline model and computed the average Spearman correlation between synthetic and real data (Table 5). The highest agreement was achieved at $A = 0.05$, which we adopted as the optimal setting. At this value, all correlations are above 0.7 and the average correlation is the highest among the tested specifications, confirming the ability of the model to capture structural patterns while allowing us to detect deviations.

***Table 5. Correlation Between Simulated and Real Data Across Attractiveness Parameters.***

| | $A$ = 0.03 | $A$ = 0.04 | $A$ = 0.05 | $A$ = 0.1 | $A$ = 0.2 |
|---|---|---|---|---|---|
| **Pioneer** | 0.760 | 0.763 | 0.759 | 0.756 | 0.749 |
| **Pioneer Impact** | 0.695 | 0.696 | 0.709 | 0.733 | 0.725 |
| **Maverick** | 0.754 | 0.746 | 0.754 | 0.740 | 0.709 |
| **Maverick Enhancement** | 0.920 | 0.918 | 0.920 | 0.925 | 0.923 |
| **Maverick Diminishment** | 0.840 | 0.840 | 0.839 | 0.838 | 0.841 |
| **Vanguard** | 0.692 | 0.672 | 0.702 | 0.689 | 0.679 |
| **Vanguard Impact** | 0.891 | 0.894 | 0.891 | 0.890 | 0.893 |
| Average | 0.7931 | 0.7899 | 0.7963 | 0.7959 | 0.7884 |

**Standardization Procedure.** We then standardized the real novelty scores using z-scores based on the simulated baseline distribution. For each real paper $r$, we computed a standardized score as $z_r = (n_r - \mu_r)/\sigma_r$, where $n_r$ is the raw novelty score of the real paper, and $\mu_r$ and $\sigma_r$ are the mean and standard deviation of the simulated novelty scores for synthetic papers with similar properties. This process highlighted deviation from expected novelty, thereby emphasizing contributions that exceeded structural randomness. To validate this standardization, we compared the aggregate distributions of real and simulated novelty scores. As shown in Figure S2, the baseline model closely approximates the empirical distribution of real data.

**Code Availability / Python Package.** The Triadic Novelty measures and their normalized versions are available through an open-source Python package named "*triadic-novelty.*" The package can be installed using *pip* or accessed directly from the GitHub repository at http://github.com/philanthrophysics/triadic-novelty.

**Data Availability:** The Web of Science Raw Data (2022 Version) employed in this paper was obtained from the Web of Science under a specific institutional agreement between Clarivate Analytics and Indiana University, which forbids the authors from sharing data derivatives. We, however, share the calculated novelty scores associated with each Web of Science ID, available at https://doi.org/10.5281/zenodo.15741054.

**Acknowledgments:** This work used JetStream2 at Indiana University through allocation CIS230183 from the Advanced Cyberinfrastructure Coordination Ecosystem: Services & Support (ACCESS) program.

**Funding**: National Science Foundation grant #2138259, #2138286, #2138307, #2137603, and #2138296. Air Force Office of Scientific Research grant #FA9550-19-1-039.

**Author contributions:** J.A., R.S.S., C.G., and F.N.S. jointly conceptualized the study and developed the methodology. J.A. and F.N.S. conducted the investigation and performed the analyses. J.A. and F.N.S. prepared the visualizations. The project was supervised by J.A., R.S.S., C.G., and F.N.S. J.A. and F.N.S. drafted the initial manuscript, and all authors contributed to reviewing and editing the final version.

**Competing interests**: All other authors declare they have no competing interests.

**Materials & Correspondence:** Correspondence and requests for materials should be addressed to J.A.

# Supplementary Information

**Database Assembling Process**

We combined journal-based and keyword-based searches to build the analysis dataset. All articles appearing in the three core journals are included. Articles published elsewhere were included based on a curated list of keywords. A detailed justification for the approach can be found in reference (36). Lists of keywords and journals are summarized below.

**List of Curated Keywords**

"philanthropy", "philanthropic", "philanthropist", "misanthropy", "misanthropic", "charity", "charities", "charitable", "benevolence", "benevolent societ*", "altruism", "altruistic", "almsgiving", "gospel of wealth", "gift exchange*", "gift-exchange*", "gifting", "gift giving", "gift-giving", "collective good*", "public good*", "the commons*", "social economy", "social economies", "gift economy", "gift economies", "grant econom*", "grant economies", "generosity", ("generous", "social"), ("generous", "people"), ("generous", "behavio*"), ("generous", "act*"), ("generous", "giving"), ("gratitude", "giving"), ("gratitude", "help*"), ("gratitude", "donate"), ("gratitude", "donation"), ("gratitude", "donating"), ("social capital", "civil societ*"), ("trust", "civil societ*"), ("communit*", "civil societ*"), "the third sector*", "voluntary sector*", "non-distribution constraint*", "nondistribution constraint*", "501c3", "501(c)(3)", "501c4", "501(c)(4)", "eleemosynary institution*","eleemosynary corporate*", "eleemosynary corporation*", "eleemosynary organi*ation*", "civic participation*", "civic engagement*", "social movement*", ("collective action*", "social"),("collective action*", "event*"), ("collective action*", "group"),("collective action*", "global"), ("collective action*", "civil societ*"), "grassroots movement*", "grassroots group*", "grassroots organi*ation*", "grassroots association*", "nonprofit*", "non-profit*", "not for profit*", "not-for-profit*", "nongovernmental organi*ation*", "nongovernmental agenc*", "non-governmental organi*ation*", "non-governmental agenc*", "nongovernmental institution*", "non-governmental institution*", "grantmaking", "grant-making", "grantseeking", "grant-seeking", "endowment foundation*", "endowment fund*", "community foundation*", "family foundation*", "private foundation*", "corporate foundation*", "social enterprise*", "social entrepreneur*", "low-profit limited liability compan*", "b corp", "benefit corporation*", "flexible purpose corporation*", "community interest compan*", "community organizing", "community engagement", "community based organi*tion*", "community-based organi*tion*", "voluntary association*", "voluntary organi*ation*", "selfhelp group*", "self-help group*", "selfhelp association*", "self-help association*", "selfhelp organi*ation*", "self-help organi*ation*", "membership association*", "membership organi*ation*", "giving circle*", ("giving", "pledge"), "donor advised fund*", "donor-advised fund*", "corporation giving", "corporate giving", "workplace giving", "impact invest*", "program*related investment*", "volunteering behavio*", "volunteerism", "voluntarism", "voluntaristics", "benefit* of volunteering", "helping other*", "helping behavio*", "giving behavio*", "individual giving", "giving money", "giving time", ("donate", "money"), ("donate", "time", "volunteer*"), ("donate", "motive*"), ("donate", "motivation*"), ("donate", "behavio*"), ("donate", "decision*"), ("donation", "money"), ("donation", "time", "volunteer*"), ("donation", "motive*"), ("donation", "motivation*"), ("donation", "behavio*"), ("donation", "decision*"), "donor fatigue", "anonymous giving", ("giving", "diaspora"), "planned giving", ("giving", "in-kind"), ("gift", "in-kind"), ("giving", "wealthy"), ("benefit* of giving"), ("indian", "giving"), ("indigenous", "giving"), ("native american", "giving"),("native-american", "giving"), "mutual aid*", ("social", "reciprocity"), "serial reciprocity", ("norm", "reciprocity"), ("norms", "reciprocity"), ("cooperation", "reciprocity"), ("generalized reciprocity", "evolution"), ("cooperative behavio*", "experiment*"), ("cooperative behavio*", "group*"),("cooperative behavio*", "human"), ("cooperative behavio*", "animal*"), ("cooperative behavio*", "social"), ("cooperation", "social"), ("cooperation", "human"), ("cooperation", "animal"), "prosocial*", "pro-social*", "empathic behavio*", "empathetic behavio*", "other regarding", "other-regarding", ("giving", "religio*"), ("donation", "religio*"), ("giving", "faith"), ("donation", "faith"), ("giving", "spiritual*"), ("donation", "spiritual*"), ("giving", "christian"), ("muslim", "giving"), ("islam", "giving"), ("buddhist", "giving"), ("jewish", "giving"), "zakat", "tzedakah", "tithe", "waqf"

**List of Keyword Combinations to Remove False Positives**
("altruistic", "suicide"), ("civil societ*", "revolution*"),("warm*glow", "familiarity"), ("charity", "mount"),("charity", "salmonella"),("charit*", "hospital*"), ("altruism", "autonomous vehicle"),("altruism", "autonomous driving"), ("charit*", "chronic")

**List of Journals**
Nonprofit and Voluntary Sector Quarterly (NVSQ, previously named Journal of Voluntary Action Research); Nonprofit Management and Leadership (NML); VOLUNTAS: International Journal of Voluntary and Nonprofit Organizations .

**Fig. S1. Descriptive Statistics of Regression Variables**

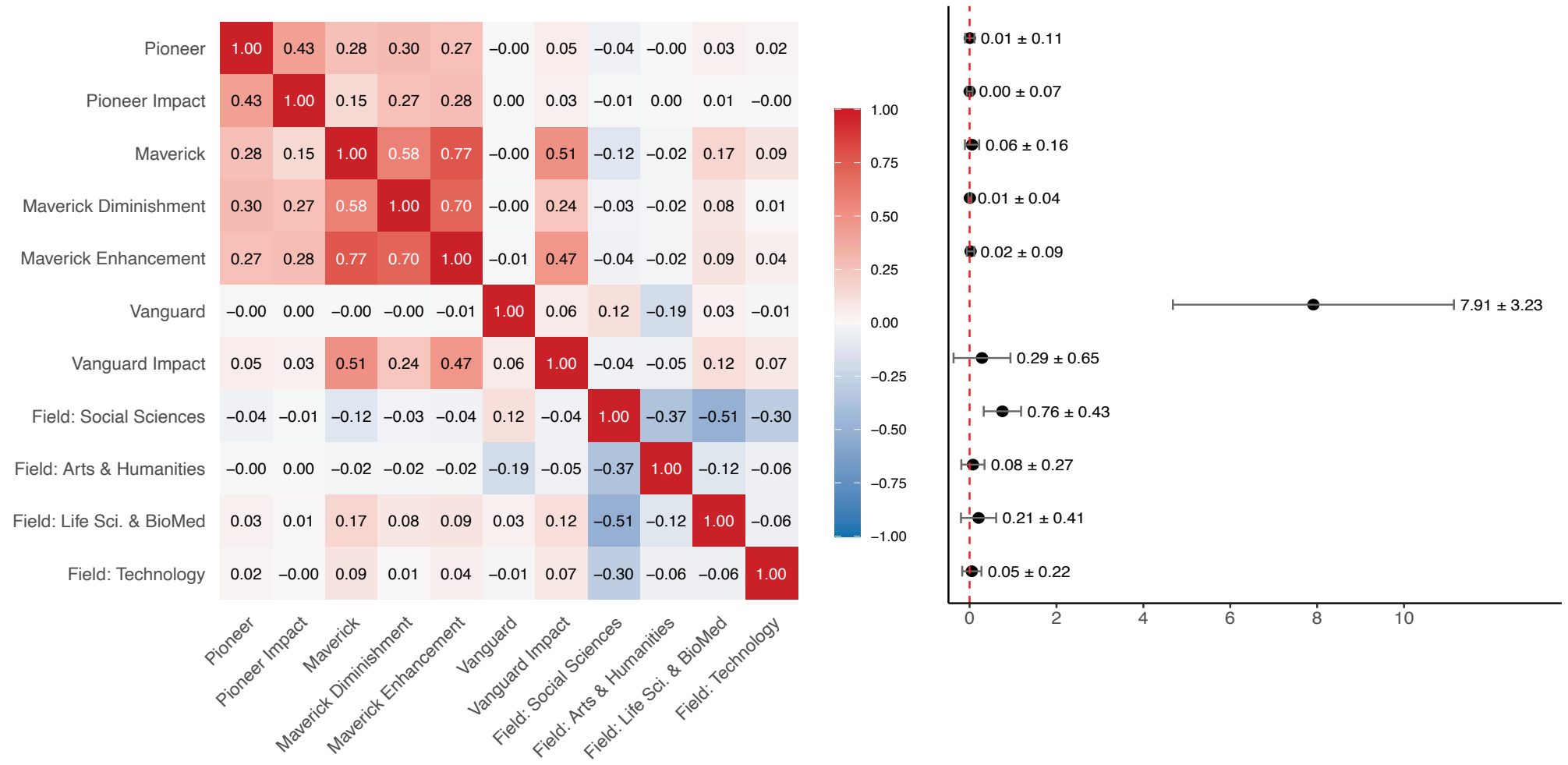

The left panel displays a Pearson correlation matrix among the variables included in the regression model. Color intensity indicates the strength and direction of the correlation, ranging from +1 (red, strong positive correlation) to -1 (blue, strong negative correlation). Diagonal values are 1.00 by definition. The right panel presents the mean and standard deviation (SD) for each variable used in the regression model. Dots indicate mean values, horizontal bars represent ±1 SD, and text labels report the summary statistics in “mean ± SD” format.

# Fig. S2. Comparison of Real and Simulated Novelty Score Distributions.

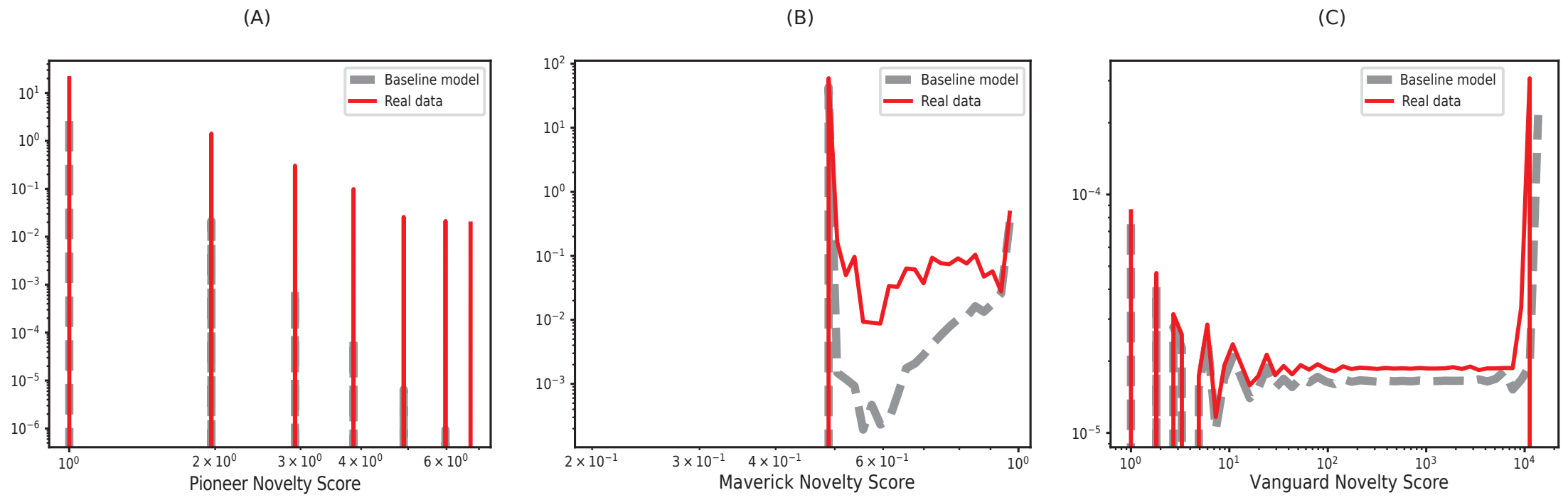